\documentclass[aps,prl,amsmath,amssymb,superscriptaddress,showpacs,10pt,reprint]{revtex4-1}

\usepackage[dvips]{graphicx}
\usepackage[colorlinks=true]{hyperref}
\usepackage[normalem]{ulem}

\graphicspath{{Figures/}}
\hypersetup{pdfpagemode = UseNone}
\usepackage{multirow}

\newcommand{\setup}[1]{{\textsf{#1}}}

\newcommand{\Id}{{{I}}}

\newcommand{\Tr}{\mathrm{Tr}}
\newcommand{\bra}[1]{\left\langle#1\right|}
\newcommand{\ket}[1]{\left|#1\right\rangle}

\newcommand{\rhoML}{\rho_\textrm{ML}}
\newcommand{\rhoTH}{\rho_\textrm{TH}}

\newcommand{\NH}{\mathcal{N}_\textrm{H}}
\newcommand{\R}{R}

\newcommand{\MODIF}[1]{{#1}}
\newcommand{\MODIFF}[1]{{#1}}
\newcommand{\DELETE}[1]{}

\begin{document}
\title{Benchmarking maximum-likelihood state estimation\\ with an entangled two-cavity state}

\author{V.~M\'etillon}
\author{S.~Gerlich}
\altaffiliation[Current affiliation: ]{Faculty of Physics, University of Vienna, Boltzmanngasse 5, A-1090 Vienna, Austria }
\author{M.~Brune}
\author{J.M.~Raimond}
\affiliation{Laboratoire Kastler Brossel, Coll\`ege de France, CNRS, ENS-Universit\'e PSL, Sorbonne Universit\'{e}, 11 place Marcelin Berthelot, F-75231 Paris, France}
\author{P.~Rouchon}
\affiliation{Centre Automatique et Syst\`{e}mes, Mines-ParisTech, PSL Research University, 60 Boulevard Saint-Michel, 75006 Paris, France}
\affiliation{INRIA Paris, 2 rue Simone Iff, 75012 Paris, France}
\author{I.~Dotsenko}
\affiliation{Laboratoire Kastler Brossel, Coll\`ege de France, CNRS, ENS-Universit\'e PSL, Sorbonne Universit\'{e}, 11 place Marcelin Berthelot, F-75231 Paris, France}
\email{igor.dotsenko@lkb.ens.fr}
\date{\today}

\begin{abstract}

The efficient quantum state reconstruction algorithm described in [P.~Six \textit{et al.}, Phys.~Rev.~A \textbf{93}, 012109 (2016)] is experimentally implemented on the non-local state of two microwave cavities entangled by a circular Rydberg atom. We use information provided by long sequences of measurements performed by resonant and dispersive probe atoms over time scales involving the system decoherence. Moreover, we benefit from the consolidation, in the same reconstruction, of different measurement protocols providing complementary information. Finally, we obtain realistic error bars for the matrix elements of the reconstructed density operator. These results demonstrate the pertinence and precision of the method, directly applicable to any complex quantum system.

\end{abstract}

\maketitle

Quantum state reconstruction or tomography is an essential operation in quantum science. It has a key role in parameter estimation, quantum metrology and studies of decoherence. It is instrumental for quantum process tomography, central in the benchmarking of quantum technologies. Many reconstruction methods have been proposed \cite{QSE, IntroQSE, Gross10, Kyrillidis18, Qi13, Torlai18,Teo11,Silberfarb05} and implemented \cite{Haeffner05, Gao10, Giovannini13, Bent15, Schoelkopf16, Riofrio17, Steffens17, Ahn19,Lvovsky01,Zambra05}.

The maximum likelihood (ML) estimation is widely used. It requires a large number of realizations of the state (density operator $\rho$). Generally, one performs a single instantaneous measurement described by a positive operator-valued measure on each of them. An iterative algorithm determines the ML estimate, $\rhoML$, of $\rho$ maximizing the likelihood of observed experimental results~\cite{Hradil97,QSE}. In principle, more information could be obtained through a composite sequence of measurements (intertwined with system evolution and decoherence), with a possibly different set of measurements for each realization. At each step of the standard ML iteration, and for each realization one then must compute the \MODIF{time} evolution through the  \MODIF{measurement} sequence to get the updated likelihood, making this procedure numerically heavy.

A recently proposed ML implementation method~\cite{Six16}, inspired by the `past quantum state' formalism~\cite{Gammelmark13}, overcomes efficiently this difficulty. The complete sequence for each realization, including information on experimental imperfections, is encapsulated in an `effect matrix', computed only once for all ML iterations. Moreover, a unique feature of this method is that it provides a direct estimate of the precision of the reconstructed density operator matrix elements. This result leads to a simple practical approach to the important problem of estimating the reconstruction precision~\cite{Christandl12,Flammia12,Blume12,Faist16}. 

In this Letter, we experimentally benchmark the power, pertinence and precision of this new method applied to the non-local state of two fields stored in two superconducting microwave cavities. The entangled state is prepared and probed by individual circular Rydberg atoms interacting sequentially with these two fields. We efficiently reconstruct the two-cavity state in a large Hilbert space by combining the results of different types of atom-cavity interactions, by taking into account imperfections and decoherence and by using all information from long sequences of probe atoms \MODIF{\DELETE{, a unique capability of the method}}.

Before turning to the experiment, let us recall briefly the main results of~\cite{Six16}. The probability $p_r(\rho)$ for observing the measurement outcomes of realization $r$ in the unknown state $\rho$ reads
	\begin{equation}\label{eq:pr}
		p_r(\rho) = \Tr \Big[ {\mathbb{K}}^{(r)}_{{N_r}} \circ {\mathbb{K}}^{(r)}_{{N_r}-1} \circ ... {\mathbb{K}}^{(r)}_2 \circ 	 {\mathbb{K}}^{(r)}_{1} (\rho) \Big],
	\end{equation}
where $\{\mathbb{K}^{(r)}_j\}$ is a sequence of $N_{r}$ time-ordered quantum maps \MODIF{\DELETE{representing all steps in the} taking into account all effects in the specific} measurement sequence: \MODIF{\DELETE{generalized measurements}} \MODIF{measurement backaction}, unitary evolutions, relaxation, etc. \MODIF{\DELETE{The probability of all results} Knowing $\rho$, the probability of all measurement records } in $\R$ independent realizations is the likelihood function $\mathcal{P}(\MODIF{\rho}) = \prod ^\R_{r=1} p_r(\MODIF{\rho})$, which is maximized by $\rhoML$.

The key ingredient in~\cite{Six16} is to write $p_r(\rho) = c_r \Tr [\rho E^{(r)}]$, 
where the effect matrix $ E^{(r)}$ reads
	\begin{equation}\label{eq:Es}
		E^{(r)} = \tilde {\mathbb{K}}^{(r)}_1 \circ \tilde {\mathbb{K}}^{(r)}_2 \circ ... \tilde {\mathbb{K}}^{(r)}_{{N_r}-1} \circ 	\tilde {\mathbb{K}}^{(r)}_{{N_r}} (\Id/ \NH).
	\end{equation}
The constant $c_r$ is a function of the measurement outcomes in realization $r$. It does not depend on $\rho$ and is thus irrelevant for the ML optimization. The `adjoint' maps, $\{ \tilde{\mathbb{K}}^{(r)}_j\}$ \cite{SM}, are applied in time-reversed order to a normalized identity operator $\Id$ ($\NH$ is a Hilbert space dimension). The effect matrix is, in the past quantum state picture~\cite{Gammelmark13}, the best estimate of the initial density matrix in realization $r$~\cite{Rybarczyk15}. Since the $E^{(r)}$s do not depend on $\rho$, they can be computed once before ML optimization, making the latter efficient even for complex sequences. Notably, the method can efficiently consolidate the results of different arbitrary measurement types and sequences. All measurements contribute to the reconstruction, even if they are performed only in a small number of realizations.

The method~\cite{Six16} also provides the confidence interval $\sigma(\langle A \rangle)$ for any observable mean value $\langle A\rangle=\Tr[A\rhoML]$~\cite{SM}. With $A=(\ket{p}\!\bra{q}\!+\!\ket{q}\!\bra{p})/2$ or $A=i(\ket{p}\!\bra{q}\!-\!\ket{q}\!\bra{p})/2$ ($\{\ket p\}$ is a Hilbert space basis), we get directly the error bars of the real and imaginary parts of $(\rhoML)_{pq}$. This direct estimation of the reconstruction quality, much simpler than other proposals~\cite{Christandl12,Flammia12,Blume12,Faist16}, makes this method particularly appealing.

\begin{figure}[t]
 		\includegraphics[width=\columnwidth]{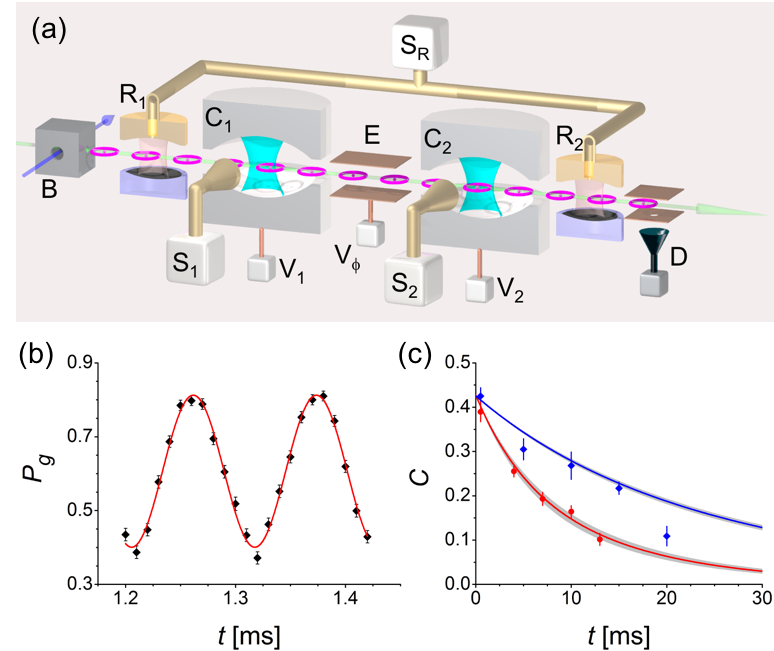}
 		\caption{Preparation and detection of a two-cavity entangled state. (a) Scheme of the experimental setup with two microwave cavities (\setup{C}$_1$, \setup{C}$_2$) and circular Rydberg atoms, see text for details. (b) Quantum beat note between the two components of the non-local state $\ket{\Psi(t)}$ directly measured as $P_g(t)$. Error bars are statistical, the solid line is a fit with $f(x) = y_0 - C/2\cos(\delta x\!+\!\varphi)$ ($0\le C\le 1$), leading to a $C_0=0.4$ contrast. (c) Time decay of $C$ revealing the decoherence of the two-cavity state. The experimental points are obtained by fitting the $P_g(t)$ oscillation signals around different $t$s for two cavity temperatures: 0.8~K (blue) and 1.5~K (red). The solid lines result from numerical predictions with $C_0$ as the only adjustable parameter. The shaded areas are numerical uncertainties resulting from the finite accuracy ($\pm 1$~ms) on the measured cavity lifetimes. \MODIFF{The residual variance between the curves and the ensemble of points is compatible with the  error bars, except for the longest time $t=20$~ms.} \MODIF{The measured frequency drift of the two-cavity setup of several tens of Hz per hour (approximate acquisition time for the measurement with $t$ = 20 ms) results in a phase diffusion significantly reducing the oscillation contrast at this timescale.} }
 		\label{fig:setup}
	\end{figure}

The experimental set-up is depicted in Fig.~\ref{fig:setup}(a). The two cavities, \setup{C}$_1$ and \setup{C}$_2$, have resonance frequencies $\omega_{c,1}/2\pi\approx\omega_{c,2}/2\pi\approx 51$~GHz. Their lifetimes are $T_{c,1}\!=\!10$~ms and $T_{c,2}\!=\!25$~ms at a 1.5~K temperature, with an average thermal photon number $n_\textrm{th}=0.25$. At 0.8~K, we have $T_{c,1}\!=\!20$~ms, $T_{c,2}\!=\!50$~ms and $n_\textrm{th}=0.06$. Sources \setup{S}$_1$ and \setup{S}$_2$ can inject coherent fields with controllable complex amplitudes in \setup{C}$_1$ and \setup{C}$_2$. The cavities states are manipulated by a sequence of individual circular rubidium Rydberg atoms (atomic states $\ket g$ and $\ket e$ with principal quantum numbers 50 and 51, atomic resonance frequency $\omega_a/2\pi=51$~GHz, atomic lifetime $T_a\approx 30$~ms)  \cite{Gleyzes07,Guerlin07}. Samples with 0.1 to 0.2 atoms on the average are \MODIF{\DELETE{excited into} prepared in} $\ket g$ in \MODIF{the excitation zone} \setup{B} out of a velocity-selected thermal atomic beam (flight time between the cavities $0.36$~ms). The common vacuum Rabi frequency measuring the atom-cavity coupling is $\Omega_0/2\pi = 49$~kHz. Applying an electric field across the mirrors of \setup{C}$_i$ with voltages sources \setup{V}$_i$, we can tune \MODIF{ \DELETE{through the Stark effect $\omega_a$ relative to $\omega_{c,i}$} $\omega_a$ relative to $\omega_{c,i}$ through the Stark effect, quadratically shifting atomic circular levels,} and thus switch between resonant ($\delta_{ac} = \omega_a-\omega_c = 0$) and dispersive ($\vert\delta_{ac}\vert > \Omega_0$) interactions and control the atom-cavity interaction time. The Ramsey zones \setup{R}$_1$ and \setup{R}$_2$, fed by source \setup{S$_\textrm{R}$}, are used to manipulate the atomic state with classical microwave pulses resonant on the $\ket g \rightarrow \ket e$ transition. The atoms are finally measured in the detector \setup{D} by state-selective field-ionisation (detection efficiency $\approx 0.5$).

We apply the reconstruction to the non-local entangled state $\ket\Psi=(\ket{1,0}+\ket{0,1})/\sqrt{2}$, where one photon is coherently shared by \setup{C}$_1$ and \setup{C}$_2$. Its preparation is reminiscent of that of an entangled state of two modes of the same cavity~\cite{Arno01}. An `entangling' atom, \setup{A}$_1$, is prepared in $\ket e$ in \setup{R}$_1$ and tuned to resonance with the initially empty cavities. It experiences a $\pi/2$-Rabi rotation in \setup{C}$_1$ and a state-swap with \setup{C}$_2$ ($\pi$-Rabi rotation). 
Due to the $\delta = \omega_{c,2}-\omega_{c,1}=2\pi\times 8.9$~kHz detuning between \setup{C}$_1$ and \setup{C}$_2$, the state evolves as $\ket{\Psi(t)} = (\ket{1,0}+e^{i\delta t}\ket{0,1})/\sqrt{2}$ within a proper phase reference and with the time origin $t=0$ set at the end of the $\pi$-Rabi pulse in \setup{C}$_2$. Because of the probabilistic Poisson distribution of atoms in a sample and their non-ideal detection, the state $\ket\Psi$ is considered to be prepared if we detect only one atom in $\ket g$. The probability to have another non-detected atom during the entanglement preparation is about $3\%$.

	 \begin{figure*}[t]
 		\includegraphics[width=\textwidth]{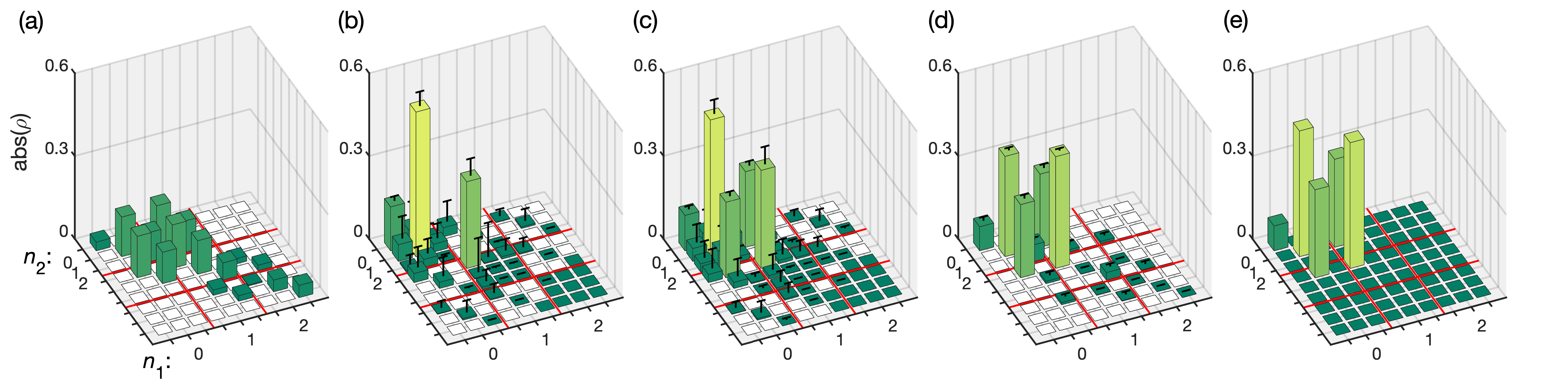}
 		\caption{Absolute values of density matrix elements in a $3\!\times\!3$ two-cavity Hilbert space. Red guiding lines enclose areas with the same photon number in \setup{C}$_1$; $n_1$ and $n_2$ are photon numbers in \setup{C}$_1$ and \setup{C}$_2$, respectively. (a)~Reconstruction with single resonant atom measurements. (b)~Reconstruction with QND parity measurement. (c) Reconstruction with both resonant and dispersive data of (a) and (b). (d)~Reconstruction with sequences of several resonant measurements. Error bars $\sigma(\textrm{abs}(\rhoML))$ in (a) are of the order of 0.5 and are not shown. Error bars smaller than $10^{-3}$ are not shown in plots (b)-(d). (e)~Numerical prediction $\textrm{abs}(\rhoTH)$ of the prepared state. The exact values of all reconstructed density matrices in the full $5\!\times\!5$ Hilbert space, together with their real and complex error bars, are given in~\cite{SM}.}
 		\label{fig:rhos}
	\end{figure*}

For an elementary check of the state preparation, we use a probe atom, \setup{A$_2$}, undoing the action of \setup{A$_1$} \cite{Arno01}. Initially in $\ket{g}$, \setup{A$_2$} performs a $\pi$-Rabi rotation in \setup{C}$_1$ and a $\pi/2$-Rabi rotation in \setup{C}$_2$. The probability $P_g$ for detecting \setup{A$_2$} in $\ket g$ at a delay time $t$ after \setup{A$_1$}  is ideally $P_g=[1- \cos(\delta t+\varphi)]/2$ ($\varphi$ is a constant phase determined by the timing details). The oscillations of $P_g$ for short times $t$ around 1.3~ms are presented in Fig.~\ref{fig:setup}(b). They have a finite contrast $C=C_0=0.4$, due to experimental imperfections. For larger times, $C$ decays with $t$, as shown in Fig.~\ref{fig:setup}(c), due to photon loss at a rate depending on the cavities temperature, 1.5~K (0.8~K) for the red (blue) dots. The solid curves are numerical predictions with $C_0$ as the only adjustable parameter.

We now proceed to a ML reconstruction of the two-cavity state. \MODIF{\DELETE{Atoms can perform resonant or dispersive measurements} We measure it by choosing between two types of the atom-cavity interaction: dispersive and resonant}. We set resonant atoms\MODIF{, prepared in $\ket g$,} to undergo the same temporal sequence as \setup{A$_2$}. In addition to the unitary evolutions and relaxation in \setup{C}$_1$ and \setup{C}$_2$, the quantum map includes the detection imperfections as well as the small probability that a second, spurious atom present in the probe sample has escaped detection~\cite{SM}. Dispersive atoms are set to implement a quantum non-demolition (QND) measurement of the joint photon-number parity of the two cavities by setting $\delta_{ac}/2\pi\approx 60$~kHz \cite{Gleyzes07}. An atomic coherence, prepared by a $\pi/2$ pulse in \setup{R}$_1$, is shifted by $\pi$ for each photon present in either cavity. This shift is probed by a $\pi/2$ pulse in \setup{R}$_2$, \MODIF{the phase of which is} set to maximize the probability for detecting the atom in $\ket g$ when both cavities are in the vacuum state. These dispersive probes do not change the photon numbers in \setup{C}$_1$ and \setup{C}$_2$. The measurements are intertwined with maps representing the free \MODIF{\DELETE{evolution} rotation} of the two-cavity state at \MODIF{the frequency difference} $\delta$ and the cavities relaxation. Optional field displacements are performed by \setup{S}$_1$ and \setup{S}$_2$ before the measurement. All quantum maps, including measurement imperfections, are given in~\cite{SM}.

We first analyse all data obtained with a single resonant probe that led to the results of Fig.~\ref{fig:setup}(b) at 0.8~K. The sequence involves a free evolution and relaxation during time $t$ followed by a single resonant probe. The reconstruction is based on $3913$ realizations and performed in a tensor Hilbert space of dimension $5\!\times\!5$ (photon numbers $n$ from 0 to 4 in each cavity). The ML estimate, $\rhoML$, is presented in Fig.~\ref{fig:rhos}(a). For the sake of visibility, we plot only the absolute values of the $\rhoML$ elements for $n$ from 0 to 2 (the full $\rhoML$ is shown in~\cite{SM}). 

Note that all two-cavity states of the form $\ket{\Xi}=(\vert n_1,n_2\rangle + \vert n_1\!-\!1,n_2\!+\!1\rangle)/\sqrt{2}$ lead, for a resonant probe, to oscillations of $P_g(t)$ with the same frequency as those produced by $\ket{\Psi(t)}$\MODIF{, since their components also differ by only one photon and thus their energy difference is also $\hbar \delta$}. The reconstructed $\rhoML$ is thus a mixture of entangled states $\ket{\Xi}$. They appear with different weights, since the dependence of the Rabi oscillation frequency on the photon number brings ambiguous information on it. Due to this photon-number indetermination, the reconstruction error bars [not shown in Fig.~\ref{fig:rhos}(a)] are extremely large, of the order of 0.5 for all elements.

It is also important to note that the ML reconstruction may be blind to some elements of $\rho$. Writing $p_r(\rho) = \sum_{p,q} (\rho_{pq})^* E^{(r)}_{pq}$ in a generic basis $\{\ket p\}$, we see that $p_r(\rho)$ does not depend on $\rho_{pq}$ if $E^{(r)}_{pq}=0$ for specific $p$ and $q$ values. More generally, if \MODIF{no measurement contains information on $\rho_{pq}$ and thus} $\sum_r |E^{(r)}_{pq}|=0$, the likelihood is independent of $\rho_{pq}$ and we get no more information on this specific matrix element than that provided by the positivity and unit trace of $\rho$. The blank elements in Fig.~\ref{fig:rhos}(a) correspond to those, on which the set of effect matrices provides no information.

An alternative measurement strategy providing better photon-number discrimination is based on QND joint parity measurements following adjustable coherent field injections (amplitudes $\alpha_1$ and $\alpha_2$) in the cavities~\cite{Deleglise08}. This procedure amounts to a direct determination of the two-cavity Wigner function, $(\pi^2/4) W(\alpha_1,\alpha_2)$, at one point in the four-dimensional phase-space~\cite{Schoelkopf16}. Here, for simplicity, we choose to inject in only one cavity at a time with 20 values of the injection amplitude ranging from 0 to~2. 

Figure~\ref{fig:rhos}(b) shows the reconstructed state using 12200 realizations, each with 40 dispersive atom samples sent over a 4-ms time period. We now fully benefit from the efficiency of the method \cite{Six16} for a long sequence of successive measurements in a single realization. The reconstruction is sensitive to the photon number (diagonal elements) and to local, single-mode coherences between states $\vert n\rangle$ and $\vert n'\rangle$ of the same cavity. Hence, the reconstructed state mainly includes $\vert 1,0\rangle \langle1,0\vert$ and $\vert 0,1\rangle\langle0,1\vert$. The significant contribution of $\vert 0,0\rangle \langle0,0\vert$ is due to atom and cavity relaxation during the state preparation. All other elements of $\rhoML$ on which we get information are zero within their error bars. Note that this measurement does not provide any information on non-local coherences between the two cavities (blank elements in the figure).

The resonant and dispersive measurements provide complementary information on $\rhoML$: the former is sensitive to non-local coherences, while the latter accurately reconstructs the photon-number probabilities. The reconstructed state consolidating the resonant and QND measurements data described above is shown in Fig.~\ref{fig:rhos}(c). Now, the dominant elements are the populations and coherences expected for $\ket\Psi$, showing that the data consolidation significantly improves the reconstruction.

As a reference to this reconstruction, Fig.~\ref{fig:rhos}(e) presents a numerical prediction, $\rhoTH$, of the prepared state. The model includes cavity and atomic relaxations, leading to the vacuum state population of 0.09. It also includes a reduction of the coherences by $30\%$ estimating the effect of stray electric fields inhomogeneity inside the atomic sample over the flight between \setup{C}$_1$ and \setup{C}$_2$. These stray fields perturb the phase of \setup{A}$_1$ and, thus, the phase between the $\ket{1,0}$ and $\ket{0,1}$ components of $\rhoTH$. They contribute to the contrast reduction observed in Fig.~\ref{fig:setup}(b). 

The method does not require that the successive measurements \MODIF{commute \DELETE{bring the minimal state perturbation}}, characteristic of the QND probes. We illustrate this unique feature by reconstructing the state with a long sequence of resonant probe samples. Each of them considerably changes the following measurement outcomes through its possible photon emission or absorption. We can nevertheless get useful information out of long sequences of non-ideally detected samples with a precise knowledge of the associated maps. The corresponding experiment involves $40$ resonant atomic samples separated by $0.2$~ms with, on \MODIF{\DELETE{the}} average $0.15$ atoms per sample. The reconstruction result based on $18000$ realizations is presented in Fig.~\ref{fig:rhos}(d). The possibility of detecting more than one atom per realization indeed considerably improves the photon number determination with respect to Fig.~\ref{fig:rhos}(a). In addition, the long measurement duration improves the discrimination of small and large $n$'s, which have different lifetimes. Finally, using many samples significantly increases the information acquisition rate per realization.


\MODIFF{We compute the fidelity $F(\rho,\sigma) = \left[\Tr \sqrt{\sqrt{\rho} \sigma \sqrt{\rho}}\right]^2$ between two different states $\rho$ and $\sigma$ in order to compare them. The fidelity of the reconstructed states of Fig.~\ref{fig:rhos}(a)-(d) with respect to the theoretical one [Fig.~\ref{fig:rhos}(e))] is 0.29, 0.85, 0.96 and 0.78, respectively. It is similar for the states (b) and (d), because the former is better in reproducing the expected populations in $\ket{1,0}$ and $\ket{0,1}$, while the former is better in the estimation of coherences.  The large fidelity value for the state in (c) highlights the interest of consolidating many different measurements in the reconstruction, a key feature of our approach.}



The method provides error bars for the density operator elements. In order to check that they faithfully describe the reconstruction precision, we apply the method to the estimation of a simple parameter and compare its predicted uncertainty to its experimental dispersion observed in many reconstructions. We have chosen to estimate the phase $\phi_a$ of the initial state $\ket{\Psi_\phi} \!=\! (\ket{1,0} + e^{i\phi_a}\ket{0,1})/\sqrt{2}$, which is an essential characteristic of the prepared entangled state. This phase is tuned by applying an electric field across the electrodes \setup{E}, sandwiched between \setup{C}$_1$ and \setup{C}$_2$ [Fig.~\ref{fig:setup}(a)]. This field changes the phase of the coherence of \setup{A}$_1$, which is finally imprinted into the phase of the two-cavity state. In the following, we prepare $\ket{\Psi_\phi}$ with $\phi_a\!=\!1.50\pm0.03$~rad, independently calibrated with Ramsey spectroscopy.

 \begin{figure}[t]
 		\includegraphics[width=1.0\columnwidth]{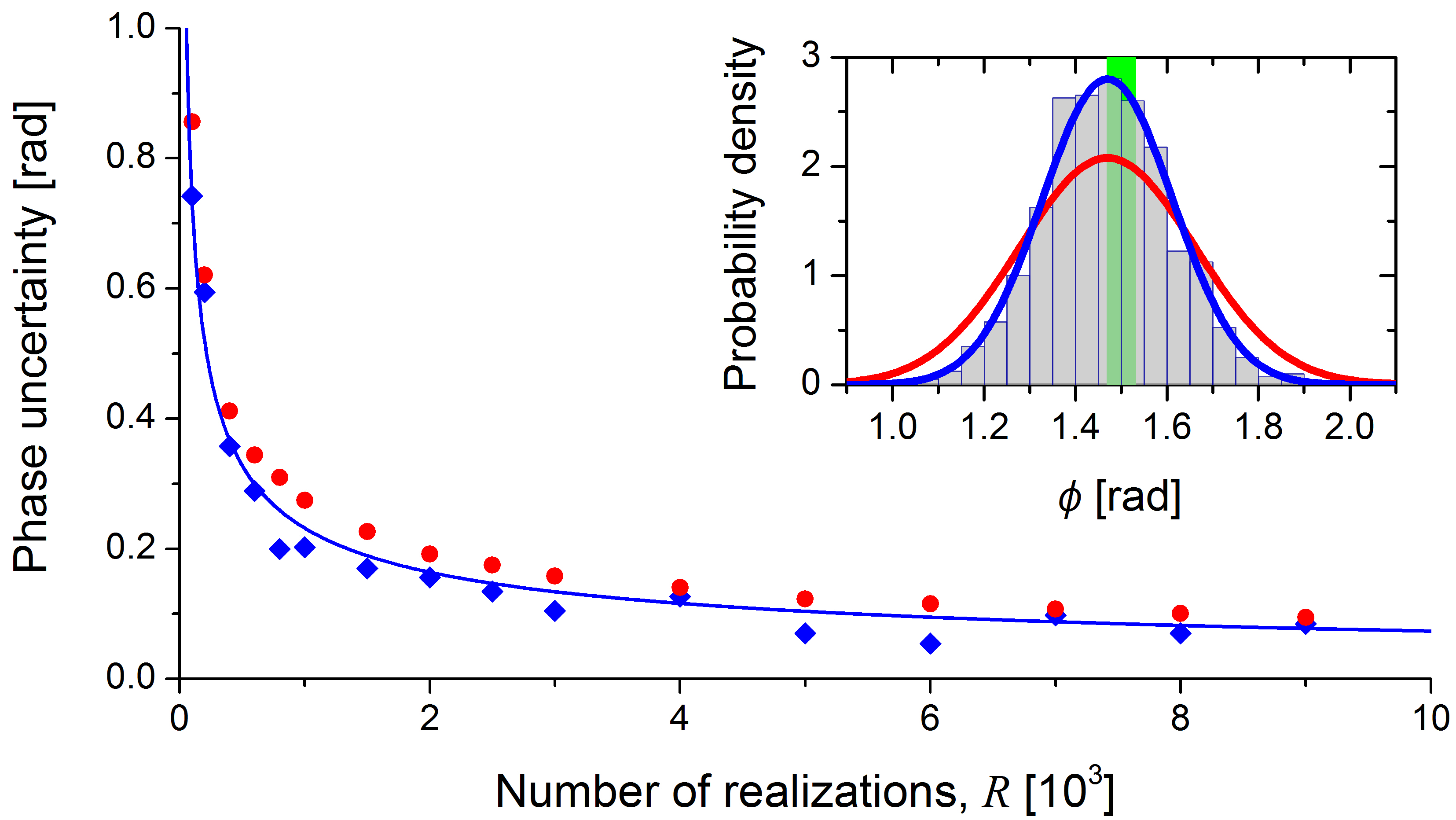}
 		\caption{Statistical properties of the phase uncertainty. 
Blue diamonds: dispersion $\tilde\sigma_{\phi,\R}$ of the reconstructed $\phi$ values versus the number of realizations $\R$. Solid line: fit with a statistical dispersion function. Red circles: mean value of computed $\sigma(\phi)$. \MODIFF{Note that having the limited set of all experimental realizations, we have fewer independent statistical sets of large size $\R$. Therefore, there are fewer independent reconstructions and hence the larger statistical variance of reconstruction results for large $\R$.} Inset: distribution of $\phi$ from 800 reconstructions on samples with $\R=2000$ realizations (histogram). Blue (red) line: Gaussian with width $\tilde\sigma_{\phi,\R}$ ($\langle\sigma(\phi)\rangle_\R$). Green band: independently measured confidence interval for $\phi_a$.}
 		\label{fig:phase}
	\end{figure}
	
\MODIF{\DELETE{The set of computed effect matrices $\{E^{(r)}\}$ is randomly split into independent groups of size $\R$ ($100\le\R\le 9000$). Using each group separately, we reconstruct $\rhoML$, determine the estimate $\phi$ of $\phi_a$ and compute its error bar $\sigma(\phi)$ \cite{SM}. With all groups of the same size $\R$, we calculate the standard deviation $\tilde\sigma_{\phi,\R}$ of the reconstructed values of $\phi$ and the mean value $\langle\sigma(\phi)\rangle_\R$ of the computed error bar~$\sigma(\phi)$. This procedure is repeated 4 times with different group samplings and the results are finally averaged.}}

\MODIF{We implement a bootstrapping approach to estimate the reconstruction precision as a function of the sample size $R$ (number of realizations used for one reconstruction), see \cite{SM} for more details. We calculate the standard deviation $\tilde\sigma_{\phi,\R}$ of the reconstructed values of $\phi$ and the mean value $\langle\sigma(\phi)\rangle_\R$ of the computed error bar~$\sigma(\phi)$.} Figure \ref{fig:phase} shows $\tilde\sigma_{\phi,\R}$ (blue diamonds) and $\langle\sigma(\phi)\rangle_\R$ (red circles) versus $\R$. Their values \MODIF{\DELETE{of $\langle\sigma(\phi)\rangle_\R$ and $\tilde\sigma_{\phi,\R}$}} are nearly equal, exhibiting the accuracy of the error bar prediction. The solid line is a fit of $\tilde\sigma_{\phi,\R}$ to a function $y(x) = A/\sqrt{x}$ confirming that the measured deviations have a purely statistical origin. The slight systematic excess of  $\langle\sigma(\phi)\rangle_R $ with respect to $\tilde\sigma_{\phi,R}$ is potentially due to  higher order correcting terms in the asymptotic expansion versus $R$ of Bayesian variances underlying  $\langle\sigma(\phi)\rangle_R$ that corresponds only  to the dominant term of order  $1/\sqrt{R}$ \cite{SixPRPralyBook2017}. The inset in Fig.~\ref{fig:phase} shows the histogram of the individual $\phi$ values for $\R=2000$. The green band gives the calibration of $\phi_a$ and its uncertainty. The blue (red) line is a Gaussian with width $\tilde\sigma_{\phi,\R}$ ($\langle\sigma(\phi)\rangle_\R$). These results confirm that the method provides realistic error bars on~$\rhoML$.

In summary, we have experimentally demonstrated on a two-cavity entangled state the pertinence and precision of the ML state reconstruction method proposed in \cite{Six16}. We have shown that it efficiently takes into account data provided by different measurement strategies as well as the system evolution during the measurement sequence. It integrates easily the description of measurement imperfections. We have shown that it provides realistic error bars for the reconstructed density matrix elements. The method is quite general and can be applied to nearly any quantum system and any measurement protocol, well beyond demonstrations in cavity or circuit QED~\cite{Six16}. \MODIF{If the initial quantum state is known, the method can also be efficiently implemented for the parameter estimation \cite{Six16a}.} Finally, the analysis of the structure of the effect matrices provides a guide for designing adaptive reconstruction procedures \cite{Qi17,Pereira18,Struchalin18} by selecting optimal measurements.\\

\begin{acknowledgments}
We acknoweldge support by European Research Council (DECLIC and TRENSCRYBE projects), by European Community (SIQS project) and by the Agence Nationale de la Recherche (QuDICE project).
\end{acknowledgments}

\end{document}


\title{Supplementary Information:\\ Benchmarking maximum-likelihood state estimation\\ with an entangled two-cavity state}

\author{V.~M\'etillon}
\author{S.~Gerlich}
\altaffiliation[Current affiliation: ]{Faculty of Physics, University of Vienna, Boltzmanngasse 5, A-1090 Vienna, Austria}
\author{M.~Brune}
\author{J.M.~Raimond}
\affiliation{Laboratoire Kastler Brossel, Coll\`ege de France, CNRS, ENS-Universit\'e PSL, Sorbonne Universit\'{e}, 11 place Marcelin Berthelot, F-75231 Paris, France}
\author{P.~Rouchon}
\affiliation{Centre Automatique et Syst\`{e}mes, Mines-ParisTech, PSL Research University, 60 Boulevard Saint-Michel, 75006 Paris, France}
\affiliation{INRIA Paris, 2 rue Simone Iff, 75012 Paris, France}
\author{I.~Dotsenko}
\affiliation{Laboratoire Kastler Brossel, Coll\`ege de France, CNRS, ENS-Universit\'e PSL, Sorbonne Universit\'{e}, 11 place Marcelin Berthelot, F-75231 Paris, France}
\email{igor.dotsenko@lkb.ens.fr}
\date{\today}

\begin{abstract}

We present in Section I all quantum maps describing the evolution of the two-cavity state in our experiment and the explicit form of all the effect matrices corresponding to the three experimental sequences described in the  main paper. In Section~II, we summarize the optimization algorithm used to find the density matrix $\rhoML$ maximizing the likelihood of the measured results. Section~III presents the evaluation of the reconstruction precision and, in particular, that of the error bars of the matrix elements of $\rhoML$. Finally, in Section~IV, we show all reconstructed density matrices in the full Hilbert space with a dimension $5\times 5$, used in the maximum likelihood optimization.

\end{abstract}

\maketitle

\section{I. State transformation maps}

Any action on a system state $\rho$ can be described by a quantum map. In this Section we present all maps corresponding to realistic, non-ideal measurements in our experiment, to cavity relaxations and to unitary transformations.

\subsection{Generalized measurement}

An unknown quantum state can be estimated by performing measurements on its many identical copies. Each copy can be exposed to a single measurement or to a series of measurements, successively projecting the initial state according to the measurement outcomes. In addition, the system can be subjected, in each realization, to state modifications due to unitary and non-unitary evolutions before and in between the measurements.

The state transformation produced  on the state  $\rho$ by a specific measurement with outcome $\mu$ can be described in terms of a Kraus operator  $M_\mu$ as  
\begin{equation}\label{eq:M_ideal}
        \mathbb{M}_\mu(\rho) = \frac{  M_\mu  \rho  M_\mu^\dag }{\Tr[M_\mu \rho M_\mu^\dag]}\ .
     \end{equation}
The product $M_\mu^\dag M_\mu$ defines a positive operator-valued measure (POVM) for outcome $\mu$. The sum of all possible POVMs for a given measurement equals to unity: $\sum_\mu M_\mu^\dag M_\mu = \Id$ \cite{HarocheBook}.


The transformation due to an unread measurement is given by
    \begin{equation}\label{eq:M_unread}
        \mathbb{M}_{\varnothing} (\rho) = \sum_\mu  M_\mu  \rho M_\mu^\dag.
    \end{equation}


\subsection{Resonant atom measurement}

Any measurement on the cavity field in our experiment is realized by a controlled atom-cavity interaction followed by atom detection. The resonant interaction with one of the cavities is described by Rabi oscillations between the atom-cavity states $\ket{g,n}$ and $\ket{e,n\!-\!1}$ with $n$ photons in the cavity.\\

\textit{One-atom coupling.}
For each initial atomic state, $\ket g$ and $\ket e$, there are two Kraus operators describing the cavity state transformation after the atom-cavity interaction and the atomic detection in  $\ket g$ or $\ket e$:
\begin{eqnarray}
	\W_{g,g}(t) &=&   \sum \cos(\chi_{n-1} t) \,\ket{n}\bra{n},\\
	\W_{g,e}(t) &=&-i \sum  \sin(\chi_{n-1} t)\, \ket{n-1}\bra{n},\nonumber\\
	\W_{e,g}(t) &=&-i \sum  \sin(\chi_{n} t)\,  	\ket{n+1}\bra{n},\nonumber\\	
	\W_{e,e}(t) &=&   \sum  \cos(\chi_{n} t)\,   	\ket{n}\bra{n},\nonumber
\end{eqnarray}
where $\chi_n = \Omega_0\sqrt{n+1}/2$ and $\Omega_0$ is the vacuum Rabi frequency of the atom-cavity interaction. The two indices of $\W$ correspond to the initial and detected atomic states, respectively.

Here, we implement two successive resonant interactions of a single atom, initially in $\ket g$, with \setup{C$_1$} and  \setup{C$_2$} during times $t_1$ and $t_2$, respectively. The Kraus operators $\Mr_{g}$ and $\Mr_{e}$, corresponding to an atomic detection in states $\ket g$ or $\ket e$ after these successive resonant interactions (hence the superscript $[res]$ standing for \textit{resonant}),  read
\begin{eqnarray}
	\Mr_{e} &=& \sum_{k} \W_{g,k}(t_1) \otimes \W_{k,e}(t_2),\\
	\Mr_{g} &=& \sum_{k} \W_{g,k}(t_1) \otimes \W_{k,g}(t_2),\nonumber
\end{eqnarray}
where the sum is taken over two intermediate (ie.~between two cavities) atomic states: $k\in\{g,e\}$. The Kraus operators are defined in the product Hilbert space of the two cavities $\{\ket{n_1} \otimes \ket{n_2}\}$. For the resonant probe atom \setup{A}$_2$ used in the main paper, we choose $ t_1=\pi/\Omega_0$ and $t_2=\pi/2\Omega_0$, corresponding to a $\pi$-Rabi rotation in \setup{C}$_1$ and to a $\pi/2$-Rabi rotation in \setup{C}$_2$.\\

\textit{Two-atom coupling.} 
The number of atoms $n_a$ per sample obeys a Poisson distribution, $P_a(n_a)$. Atomic samples may thus contain more than one atom. For instance, for a mean atom number equal to $0.1$, the probability of having one and two atoms in a sample is $P_a(1)=9.05\%$ and $P_a(2) = 0.45\%$, respectively (three-atom events can be ignored). Two-atom events have a non-negligible contribution and must be taken into account in the Kraus map construction. In general, two atoms can be prepared and detected in one of the four possible states: $|ee\rangle$, $|eg\rangle$, $|ge\rangle$, and $|gg\rangle$. Their coupling to one cavity during the interaction time $t$ is described by the Kraus operators

\small
\begingroup
\allowdisplaybreaks
\begin{eqnarray}\label{eqn_2atoms1}
  \W_{ee,ee}(t) & = & \sum \left[1+\frac{n+1}{2n+3}\!\left(\!\cos\Big(\xi_{n+1} t\right)\!-\!1\Big)\!\right]\ket{n}\bra{n},  \\
  \W_{ee,eg}(t) & = & -i\sum \sqrt{\frac{n+1}{2(2n+3)}}\sin\Big(\xi_{n+1} t\Big) \ket{n\!+\!1}\bra{n}, \nonumber \\
  \W_{ee,ge}(t) & = & \W_{ee,eg}(t), \nonumber \\
  \W_{ee,gg}(t) & = & \sum \frac{\sqrt{\left(n+1\right)\left(n\!+\!2\right)}}{2n\!+\!3}\left(\cos\Big(\xi_{n+1} t\right)-1\Big)\ket{n\!+\!2}\bra{n},\nonumber \\
  		& & \nonumber \\
  		& & \nonumber \\
  \W_{eg,ee}(t) & = & -i\sum \sqrt{\frac{n+1}{2(2n+3)}}\sin\Big(\xi_{n+1} t\Big)\ket{n}\bra{n\!+\!1}, \nonumber \\
  \W_{eg,eg}(t) & = & \frac{1}{2}\sum \left[1+\cos\Big(\xi_n t\Big)\right] \ket{n}\bra{n}, \nonumber \\
  \W_{eg,ge}(t) & = & \frac{1}{2}\sum \left[-1+\cos\Big(\xi_n t\Big)\right] \ket{n}\bra{n},  \nonumber \\
  \W_{eg,gg}(t) & = & -i\sum \sqrt{\frac{n+1}{2(2n+1)}}\sin\Big(\xi_n t\Big) \ket{n\!+\!1}\bra{n},\nonumber \\
		& & \nonumber \\
		& & \nonumber \\
  \W_{ge,ee}(t) & = & \W_{eg,ee}(t), \nonumber \\
  \W_{ge,eg}(t) & = & \W_{eg,ge}(t), \nonumber \\
  \W_{ge,ge}(t) & = & \W_{eg,eg}(t),  \nonumber \\
  \W_{ge,gg}(t) & = & \W_{eg,gg}(t), \nonumber \\
		& & \nonumber \\
  		& & \nonumber \\
  \W_{gg,ee}(t) & = & \sum \frac{\sqrt{\left(n\!+\!1\right)\left(n\!+\!2\right)}}{2n+3}\left(\cos\Big(\xi_{n+1} t\Big)-1\right) \ket{n}\bra{n\!+\!2}, \nonumber \\
  \W_{gg,eg}(t) & = & -i\sum \sqrt{\frac{n+1}{2(2n+1)}}\sin\Big(\xi_n t\Big)\ket{n}\bra{n\!+\!1}, \nonumber\\
  \W_{gg,ge}(t) & = & \W_{gg,eg}(t), \nonumber \\
  \W_{gg,gg}(t) & = & \sum \left[1+\frac{n}{2n-1}\left(\cos\Big(\xi_{n-1} t\Big)-1\right)\right] \ket{n}\bra{n}.\nonumber
\end{eqnarray}
\endgroup
\normalsize
Here, all sums start from $n=0$, $\xi_n = \Omega_0 \sqrt{n+1/2}$ and $\xi_{-1}$ is defined to be zero for the sake of notational convenience. The derivation of these transformations can be found, for instance, in Chapter~II.3 of \cite{Zhou12}. 

The successive interaction of two atoms, initially in state $|gg\rangle$, with the two cavities is described by the two-cavity Kraus operators:
\begin{eqnarray}
	\Mr_{ee} &=& \sum_{k} \W_{gg,k}(t_1) \otimes \W_{k,ee}(t_2),\\
	\Mr_{eg} &=& \sum_{k} \W_{gg,k}(t_1) \otimes \W_{k,eg}(t_2),\nonumber\\
	\Mr_{eg} &=& \Mr_{ge} ,\nonumber\\
	\Mr_{gg} &=& \sum_{k} \W_{gg,k}(t_1) \otimes \W_{k,gg}(t_2),\nonumber
\end{eqnarray}
where the sum is taken over  the four intermediate two-atom states: $k\in\{ee,eg,ge,gg\}$.

\subsection{Dispersive (QND) atom measurement}

We use the dispersive interaction in order to perform a quantum non-demolition (QND) measurement of the joint photon-number parity of the two cavities. It is implemented by the  \setup{R}$_1$--\setup{R}$_2$ Ramsey interferometer. For an atom-cavity detuning $\delta_{ac}/2\pi\approx 60$~kHz, we achieve a $\pi$-phase shift per photon for the atomic $\ket g\!-\!\ket e$ coherence \cite{Gleyzes07,Deleglise08}. The interferometer phase is set to maximize the probability for finally detecting the atom in $\ket g$ when both cavities are empty. The detection of a dispersive probe atom in $\ket g$ or $\ket e$ corresponds to the Kraus operators (the superscript $[dis]$ stands for \textit{dispersive}): 
    \begin{eqnarray}\label{eq:QND}
        \Md_{g} &=& \cos(\N\pi/2),\\
        \Md_{e} &=& \sin(\N\pi/2),\nonumber
    \end{eqnarray}
where $\N\!=\!\N_1\!\otimes\! I_2\!+\!I_1\! \otimes \!\N_2$ is the sum of  the photon-number operators $N_i$ of the two cavities ($I_i$ are identity operators).

To a good approximation the simultaneous interactions of two atoms of the same sample with the cavities can be treated separately. The state transformation after detecting two atoms in states $s_1$ and $s_2$ is given by the product of the corresponding one-atom operations: $\Md_{s1,s2} = \Md_{s1}\Md_{s2}$.

\subsection{Non-ideal atom detection}\label{ss:imperfections}

   \begin{table*}[]
    \centerline{
        \begin{tabular}{|c||c|c|c|c|c|c|}
          \hline
             $\,\mu'\setminus \mu\,$ &
             $\,\varnothing\,$ &
             $g$ &
             $e$ &
             $gg$ &
             $ee$ &
             $ge$
        \vph\\\hline\hline
             $\varnothing$ &
             $1$ &
             $1\!-\!\varepsilon$ &
             $1\!-\!\varepsilon$ &
             $(1\!-\!\varepsilon)^2$ &
             $(1\!-\!\varepsilon)^2$ &
             $(1\!-\!\varepsilon)^2$
        \vph\\\hline
             $g$ &
             $0$ &
             $\varepsilon(1\!-\!\eta_g)$ &
             $\varepsilon\eta_e$ &
             $\,2\varepsilon(1\!-\!\varepsilon)(1\!-\!\eta_g)$ &
             $\,2\varepsilon(1\!-\!\varepsilon)\eta_e$ & $\varepsilon(1\!-\!\varepsilon)(1\!-\!\eta_g\!+\!\eta_e)\,$
        \vph\\\hline
             $e$ &
             $0$ &
             $\varepsilon\eta_g$ &
             $\varepsilon(1\!-\!\eta_e)$ &
             $\,2\varepsilon(1\!-\!\varepsilon)\eta_g$ &
             $\,2\varepsilon(1\!-\!\varepsilon)(1\!-\!\eta_e)$ &   $\varepsilon(1\!-\!\varepsilon)(1\!-\!\eta_e\!+\!\eta_g)\,$
        \vph\\\hline
             $gg$ &
             $0$ &
             $0$ &
             $0$ &
             $\varepsilon^2(1\!-\!\eta_g)^2$ &
             $\varepsilon^2\eta_e^2$ &
             $\varepsilon^2\eta_e(1\!-\!\eta_g)$
        \vph\\\hline
             $ge$ &
             $0$ &
             $0$ &
             $0$ &
             $\,2\varepsilon^2\eta_g(1\!-\!\eta_g)$ &
             $\,2\varepsilon^2\eta_e(1\!-\!\eta_e)$ &   $\,\varepsilon^2((1\!-\!\eta_g)(1\!-\!\eta_e)\!+\!\eta_g\eta_e)\,$
        \vph\\\hline
             $ee$ &
             $0$ &
             $0$ &
             $0$ &
             $\varepsilon^2\eta_g^2$ &
             $\varepsilon^2(1\!-\!\eta_e)^2$ &
             $\varepsilon^2\eta_g(1\!-\!\eta_e)$
        \vph\\\hline
        \end{tabular}
    }
    \caption{Stochastic matrix showing the probability $P(\mu'|\mu)$ to measure outcome $\mu'$ for each ideal measurement outcome $\mu$.}
    \label{tb:stochastic}
    \end{table*}

Any measurement on the cavity field in our experiment is based on atom detection. The number of atoms $\na$ in each sample follows the Poisson distribution law, $\Pa(\na)$,  with the mean atom number of up to 0.2. If we ignore the probability of having more than 2 atoms in a sample, there are six possible detection results: $\mu\in\{\varnothing,g,e,gg,ee,ge\}$, where $\varnothing$ denotes no detected atom. The superoperators $\mathbb{L}_{\mu}^{[\nu]}$ describing the field evolution produced by the interaction $\nu\in\{res,dis\}$ with a sample finally measured in state $\mu$ are thus given by
    \begin{equation}
        \label{eq:L_sensor_2}
        \begin{array}{lcrl}\vspace{0.1cm}
        \mathbb{L}_{\varnothing}^{[\nu]} &=& \Pa(0)    &\mathbb{\Id},           \\\vspace{0.1cm}
        \mathbb{L}_{g}^{[\nu]}           &=& \Pa(1)    &\mathbb{M}_g^{[\nu]} ,      \\\vspace{0.1cm}
        \mathbb{L}_{e}^{[\nu]}           &=& \Pa(1)    &\mathbb{M}_e^{[\nu]},      \\\vspace{0.1cm}
        \mathbb{L}_{gg}^{[\nu]}        &=& \Pa(2)    &\mathbb{M}_{gg}^{[\nu]} ,   \\\vspace{0.1cm}
        \mathbb{L}_{ge}^{[\nu]}      &=& 2\,\Pa(2) &\mathbb{M}_{ge}^{[\nu]},   \\\vspace{0.1cm}
        \mathbb{L}_{ee}^{[\nu]}      &=& \Pa(2)    &\mathbb{M}_{ee}^{[\nu]},
        \end{array}
    \end{equation}
where $\mathbb{\Id}$ is the unity superoperator.

If the imperfect measurement apparatus is not able to completely distinguish different $\mu$'s, the real transformation is a mixture of projections of all ideal outcomes $\mu$ compatible with the measured result $\mu'$:
    \begin{equation}\label{eq:M_real}
        \mathbb{S}^{[\nu]} _{\mu'} (\rho) = \frac{\sum_\mu P(\mu'|\mu)  \,\mathbb{L}^{[\nu]} _{\mu} (\rho)}
        {\Tr[\sum_\mu P(\mu'|\mu) \,\mathbb{L}^{[\nu]} _{\mu} (\rho)]}\ .
    \end{equation}
Here, $P(\mu'|\mu)$ is the probability to obtain $\mu'$ for each~$\mu$ and the sum is taken over all possible measurement results $\mu$~\cite{Peaudecerf13}.

Our non-ideal atom detector has a reduced detection efficiency of $\varepsilon\approx0.5$, i.e.~only one atom out of two is detected on the average. In addition, with the probability of $\eta_g=0.05$ ($\eta_e=0.07$), the atomic state $\ket g$ ($\ket e$) can be erroneously detected as the opposite state  $\ket e$ ($\ket g$). The conditional probabilities $P(\mu'|\mu)$ are given as a stochastic matrix in Table~I and explained in detail in~\cite{Peaudecerf13}. 

\subsection{Cavity field relaxation}

The cavities have limited lifetimes $\Tc$ and a non-zero thermal photon number $\nth$. Therefore, the relaxation of state $\rho_i$ in cavity \setup{C}$_i$ during a short time interval $\tau=\xi\Tc$, with $\xi\ll 1$, can be approximated by the action of the superoperator 
$\mathbb{T}$:
    \begin{equation}\label{eq:rho_jump}
        \mathbb{T}_{i,\tau}(\rho_i) = \JumpO_{,\tau} \rho_i \JumpO_{,\tau}^\dag  + \JumpD_{,\tau} \rho_i \JumpD_{,\tau}^\dag + \JumpU_{,\tau} \rho_i \JumpU_{,\tau}^\dag,
    \end{equation}
where  the jump operators are defined as 
    \begin{eqnarray}
        \JumpO_{,\tau} &=& (1-\xi \nth/2)\,\Id - \xi(1/2 + \nth) \, \ba^\dag \ba,\\
       \JumpD_{,\tau} &=& \sqrt{\xi(1+\nth)} \, \ba,\\
        \JumpU_{,\tau} &=& \sqrt{\xi \nth} \, \ba^\dag,
    \end{eqnarray}
with $\Id$ the identity operator and $a$ the photon-number annihilation operator (we omit the indices $i$ when they are not essential). They describe events in which the photon number changes by $0$, $-1$ and $+1$, respectively (see, eg., \cite{Peaudecerf13}). The evolution of the joint two-cavity state is given by the map
    \begin{equation}\label{eq:rho_decoherence}
        \mathbb{T}_{\tau} (\rho) = (\mathbb{T}_{1,\tau} \otimes \mathbb{T}_{2,\tau} ) (\rho).
    \end{equation}

\subsection{Unitary transformation maps}
    
Besides the non-unitary measurement- and environment-induced evolutions, the system state $\rho$ undergoes  unitary evolutions $U$  corresponding to the generic map 
    \begin{equation}\label{eq:U}
        \mathbb{U} (\rho) =U \rho\, U^\dag.
    \end{equation}

We use coherent amplitude injections in  \setup{C}$_i$, by means of the microwave source \setup{S}$_i$,  to provide phase sensitivity to a QND measurement \cite{Deleglise08}  (this is equivalent to a homodyne measurement in quantum optics, giving access to the optical field quadratures). The field injection of a complex amplitude $\alpha$ realizes a unitary displacement  $D(\alpha) = \exp (\alpha a^\dag-\alpha^* a)$, modifying the state $\rho_i$ of cavity \setup{C}$_i$ as 
    \begin{equation}\label{eq:D}
        \mathbb{D}_{\alpha}(\rho_i) = D(\alpha) \rho_i D^\dag(\alpha).
    \end{equation}
More generally, the simultaneous injection of coherent fields $\alpha_1$ and $\alpha_2$ into two cavities is described by the Kraus map 
    \begin{equation}\label{eq:D1D2}
        \mathbb{D}_{\alpha_1,\alpha_2} (\rho) = (\mathbb{D}_{\alpha_1} \otimes \mathbb{D}_{\alpha_2} ) (\rho).
    \end{equation}

In the experiment, we detune the frequencies of the two cavities by $\delta$. Their joint state is  thus always subject to a unitary evolution, given by the operator $R(\tau) = \exp(i\delta \tau (\N_2-\N_1)/2)$ for an evolution time $\tau$. The corresponding Kraus map reads 
\begin{equation}\label{eq:rho_rotation}
      \mathbb{R}_{\tau} (\rho) =R(\tau) \rho R^\dag(\tau).
\end{equation}\\

\subsection{Adjoint maps}

The definition (2) of an effect matrix $E^{(r)}$ in the main paper involves adjoint maps, $\{ \tilde{\mathbb{K}}^{(r)}_j\}$. The adjoint map $\tilde{\mathbb{K}}$ of $\mathbb{K}$ is defined~by 
\begin{equation}\label{eq:adjoint}
     \Tr[A \,\mathbb{K}(B)] = \Tr[\tilde{\mathbb{K}}(A) B]
\end{equation}
for any Hermitian operators $A$ and $B$ \cite{Six16}. Therefore, for a map in the general form $\mathbb{K}(\rho) = \sum_i K_i \rho K^\dag_i$, its adjoint reads
\begin{equation}\label{eq:adjoint2}
    \tilde{\mathbb{K}}(\rho) = \sum_i K^\dag_i \rho K_i.
\end{equation}
This gives us a simple way for calculating all adjoint maps used in our maximum likelihood reconstruction ($\tilde {\mathbb{S}}^{[res]}$,  $\tilde {\mathbb{S}}^{[dis]}$,  $\tilde {\mathbb{T}}$, $\tilde {\mathbb{D}}$ and $\tilde {\mathbb{R}}$) via the definition of the corresponding standard maps.

\subsection{Effect matrices for three experimental sequences}

In the main paper we present three types of experimental sequences measuring the two-cavity entangled state. The simplest sequence used for the data shown in Fig.~2(a) consists of a fixed waiting time $t$ (typically, about 1~ms) and a single-resonant-atom interaction. Since decoherence (field relaxation) $\mathbb{T}$ and phase shift $\mathbb{R}$ do not commute, we compute their action over the waiting time $t$ by concatenating small time intervals $\tau=0.1$~ms, much smaller than $T_{c,1}$ and~$T_{c,2}$, in the Trotter approximation spirit. In each time interval, we approximate the evolution by sequential actions of decoherence and phase shift. As a result, the complete effect matrix reads
\begin{equation}\label{eq:E_1res}
      E_{1res}^{(r)} = \Big( \tilde{\mathbb{T}}_{\tau}\circ \tilde{\mathbb{R}}_{\tau}\Big)^{t/\tau}   \circ \tilde {\mathbb{S}}^{[res]}_{\mu'} (\Id/ \NH),
\end{equation}
where $\NH$ is the Hilbert space dimension and $\mu'$ is the atom detection outcome in realization $r$.

The measurement sequence used for Fig.~2(b) consists of a field displacement after time $t$, followed by a sequence of $\Ns$ dispersive (QND) atomic samples, separated by time $t_a$. The corresponding effect matrix reads  
\begin{eqnarray}\label{eq:E_qnd}
      E_{qnd}^{(r)} &=& \Big( \tilde{\mathbb{T}}_{\tau}\circ \tilde{\mathbb{R}}_{\tau}\Big)^{t/\tau}  \tilde{\mathbb{D}}_{\alpha_1,\alpha_2}  \circ \\
      			 & & \prod_{j=1}^{\Ns-1} \Big( \tilde {\mathbb{S}}^{[dis]}_{\mu'(j)} \circ \tilde{\mathbb{T}}_{t_a}\circ \tilde{\mathbb{R}}_{t_a}\Big)
      			\circ \tilde {\mathbb{S}}^{[dis]}_{\mu'(\Ns)} (\Id/ \NH),\nonumber
\end{eqnarray}
with $\mu'(j)$ being the detection result of the sample number~$j$ in the realization $r$.

Finally, the last experimental sequence used for Fig.~2(d) involves $\Ns$ resonant atomic samples, separated by $t_a$, send at time $t$ after the state preparation:
\begin{eqnarray}\label{eq:E_Nres}
      E_{Nres}^{(r)} &=& \Big( \tilde{\mathbb{T}}_{\tau}\circ \tilde{\mathbb{R}}_{\tau}\Big)^{t/\tau}  \circ\\
      			 & & \prod_{j=1}^{\Ns-1} \Big( \tilde {\mathbb{S}}^{[res]}_{\mu'(j)} \circ \tilde{\mathbb{T}}_{t_a}\circ \tilde{\mathbb{R}}_{t_a}\Big)
      			\circ \tilde {\mathbb{S}}^{[res]}_{\mu'(\Ns)} (\Id/ \NH).\nonumber
\end{eqnarray}\\


\section{II. Algorithm for the Maximum Likelihood optimisation}

The maximization of the likelihood function $\mathcal{P}(\varrho)$ is typically realized via the maximization of the concave log-likelihood function 
\begin{eqnarray}
	f(\rho)=\log \mathcal{P}(\varrho) = \sum ^\R_{r=1} \log \left(\Tr [\rho E^{(r)}]\right)
\end{eqnarray}
over the convex closed set of density operators $\rho$ ($R$ is the total number of measurement sequences). This maximization is  based on the iterative gradient algorithm with orthogonal projection starting from the initial guess $\rho_1$. After the step number $k$, the state estimate $\rho_{k+1}$ is calculated by
\begin{eqnarray}\label{eq:rho_k}
	\rho_{k+1}= \Pi \big(\rho_k + g_k \nabla f_k \big)
\end{eqnarray}
with  
\begin{eqnarray}
	\nabla f_k=\sum ^\R_{r=1} \frac{ E^{(r)}}{\Tr [\rho_k E^{(r)}]} ,
\end{eqnarray}
where $\Pi$ stands for  the orthogonal projection onto the set of density operators. It is defined as follows. For any Hermitian operator $X= U \Delta U^\dag$ with $U$ unitary and $\Delta$ diagonal ($\Delta=\text{diag}(\lambda)$),  $\Pi(X)= U \Pi(\Delta) U^\dag$ with $\Pi(\Delta)$ also diagonal and corresponding to the orthogonal projection of the vector $(\lambda_1,\lambda_2, \ldots)$ onto  the simplex $\left\{(p_1, p_2,\ldots)~\big|~ \forall j,~p_j \geq 0,~\sum_{j} p_j=1\right\}$. The positive parameter $g_k$ in \eqref{eq:rho_k} is chosen such that  $g_k \| \nabla^2 f_k \|$ is not too large. We set 
\begin{eqnarray}
	g_k = -\frac{\Tr[(\nabla f_k)^2]}{\nabla^2 f_k (\nabla f_k, \nabla f_k)},
\end{eqnarray}
where Hessian is given by 
\begin{eqnarray}
	\nabla ^2 f_k(X,X) = -\sum\frac{\Tr^2[E^{(r)} X]}{\Tr^2[\rho_k E^{(r)}]}.
\end{eqnarray}

Our typical initial guess is $\rho_1 = \Id/\NH$. Iterations are stopped at step $k$ when the optimality conditions, given by Eq.~(8) in~\cite{Six16}, are satisfied up to a numerical tolerance $\epsilon= 10^{-7} $: $\rhoML$ is set to $\rho_k$ when 
\begin{eqnarray}
  \big\| \rho_k \nabla f_k - \nabla f_k \, \rho_k  \big\| & \leq &\epsilon \|\nabla f_k \|,  \\
  \big\| P_k \nabla f_k - \lambda_k  P_k \big\| &\leq & \epsilon \big\| P_k \nabla f_k P_k \big\| + \epsilon \big\| \rho_k \big\|,  \\
  \lambda_{\min} \Big[ \lambda_k I - \nabla f_k  \Big] &\leq & \epsilon \big\|\lambda_k \Id\big\| + \epsilon \big\| \nabla f_k\big\|,
\end{eqnarray}
where $\| X \|=\sqrt{\Tr[X^2]}$ stands for  the Frobenius norm of the Hermitian operator $X$, $ \lambda_{\min}(X)$ for its smallest eigenvalue, $P_k$  for the orthogonal projection on the range of $\rho_k$ and $\lambda_k=\Tr[P_k \nabla f_k]/ \Tr[P_k]$ (the eigenvalues of $\rho_k$ less than $\epsilon$ are set to zero).

\section{III. Precision of the quantum state reconstruction}

For any Hermitian operator $A$,  the ML estimate of its  average value,  $\langle A\rangle=\Tr[A\rhoML]$, admits a  standard deviation  $\sigma(\langle A \rangle)$ that can be approximated by
\begin{eqnarray}
	\sigma^2(\langle A \rangle) \approx   \Tr\left[A_{\parallel} ~\boldsymbol{\mathcal{R}}^{-1}(A_\parallel)\right],
\end{eqnarray}
where
\begin{eqnarray}
	A_{\parallel}=A \!-\! \frac{\Tr[A P_\textrm{ML}]}{\Tr[P_\textrm{ML}]} P_\textrm{ML} \!-\! (I\!-\!P_\textrm{ML}) A (I\!-\!P_\textrm{ML})
\end{eqnarray}
with $P_\textrm{ML}$ the orthogonal projector on the range of $\rhoML$, see~\cite{SixPRPralyBook2017} and Eqs.~(10)--(11) in~\cite{Six16}. The linear super-operator $\boldsymbol{\mathcal{R}}$ is given by Eq.~(11) in \cite{Six16}. It is self-adjoint and non-negative for the Frobenius scalar product between Hermitian operators. When $\boldsymbol{\mathcal{R}}$  is not positive definite,  $\boldsymbol{\mathcal{R}}^{-1}$ corresponds to its Moore-Penrose pseudo-inverse. Numerically, $\boldsymbol{\mathcal{R}}^{-1}(A)$  is obtained via usual linear algebra routines just by considering $A$ as a complex vector of length $\NH^2$ (where $\NH$ is the dimension of the underlying Hilbert space) and $\boldsymbol{\mathcal{R}}$ as a $\NH^2\times \NH^2$ Hermitian matrix.

Here, we use $\sigma^2(\langle A \rangle)$ to calculate the error bars of the reconstructed density matrix $\rhoML$. The density matrix element $\rho_{pq}$ is, in general, complex and we use the following standard notations:
\begin{eqnarray}\label{eq:complex}
     x_{pq} &=& \textrm{re}(\rho_{pq}),\\
     y_{pq} &=& \textrm{im}(\rho_{pq}),\nonumber\\
     \phi_{pq} &=& \textrm{arg}(\rho_{pq}),\nonumber\\
     r_{pq} &=& \textrm{abs}(\rho_{pq}) = |\rho_{pq}|.\nonumber
\end{eqnarray}
We denote $\sigma(x_{pq})$ and $\sigma(y_{pq})$ the precision (ie.~error bars) of the real ($ x_{pq} $) and imaginary ($y_{pq}$) parts of a reconstructed density matrix element $\rho_{pq}$, respectively. These precisions are equal to $\sigma(x_{pq}) = \sigma(\langle X^{(pq)}\rangle)$  and $\sigma(y_{pq}) = \sigma(\langle Y^{(pq)}\rangle)$, calculated for the following operators:
\begin{eqnarray}\label{eq:sigmas}
     X^{(pq)} &=& \frac{\ket{p}\bra{q}+\ket{q}\bra{p}}{2},\\
     Y^{(pq)} &=& i\frac{\ket{p}\bra{q}-\ket{q}\bra{p}}{2}.\nonumber
\end{eqnarray}
Knowing the simple dependence of $\phi_{pq}$ and $r_{pq}$ on $x_{pq}$ and $y_{pq}$, we easily get the corresponding precision of the phase and the absolute value of $\rho_{pq}$:
\begin{eqnarray}\label{eq:sigmas}
     \sigma(\phi_{pq}) &=& \frac{\sqrt{\big[y_{pq}\,\sigma(x_{pq})\big]^2 \!\!+\!\big[x_{pq}\,\sigma(y_{pq})\big]^2}} {r_{pq}^2},\\
     \sigma(r_{pq})      &=& \frac{\sqrt{\big[x_{pq}\,\sigma(x_{pq})\big]^2 \!\!+\!\big[y_{pq}\,\sigma(y_{pq})\big]^2}} {r_{pq}}.\nonumber
\end{eqnarray}

\section{IV. Reconstructed density matrices}

All reconstructions presented in the paper have been realized in the Hilbert space dimension $\NH=5\times 5$. However, for the sake of clarity, in the main text of the paper we plot the reconstructed density matrices limited to 2 photons (see Fig.~2 of the main paper). Here, in Fig.~S1, we show full reconstructed density matrices of size $5^2\times5^2$. \MODIF{The fidelity of states (a)-(d)  with respect to the expected one is 0.29, 0.85, 0.96 and 0.78, respectively.} Numerical values of all presented density matrices, as well as their error bars, are given in additional files, compressed in an attached zip archive.

	  \begin{figure*}
    		\includegraphics[width=1.05\textwidth]{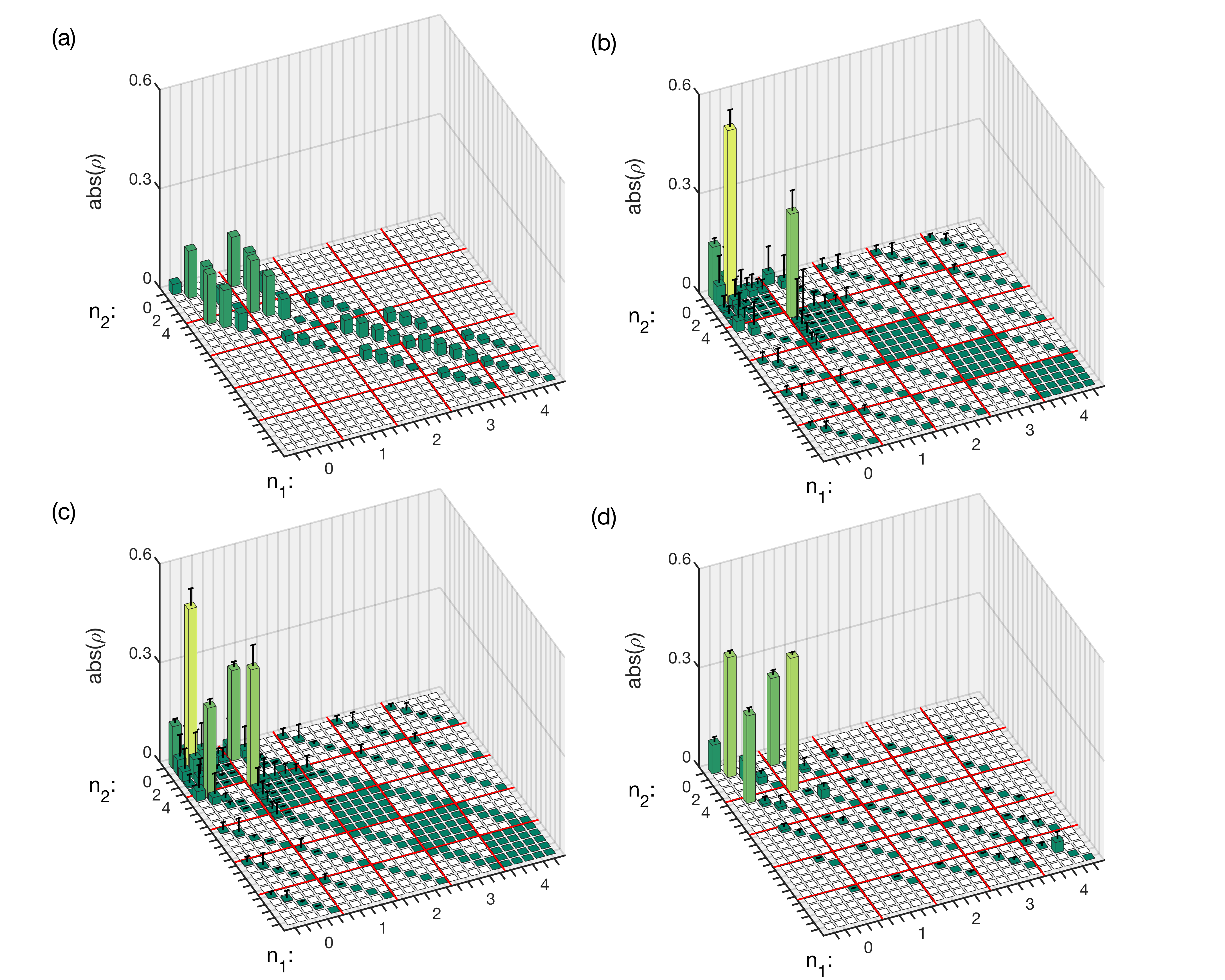}
  		\caption{Density matrix absolute values in a $5\!\times\!5$ two-cavity Hilbert space. Red guiding lines enclose areas with the same photon number in \setup{C}$_1$. Blank elements correspond to zero elements in all $E^{(r)}$, see main text for details.\\  		
  		(a)~Reconstruction based on single resonant-atom measurements.\\
  		(b)~Reconstruction based on QND parity measurement.\\
  		(c) Reconstruction with both resonant and dispersive data of (a) and (b).\\
  		(d)~Reconstruction with sequences of several resonant atoms.\\
  		Error bars $\sigma(\textrm{abs}(\rhoML))$ in (a) are of the order of 0.5 and are not shown. Error bars smaller than $10^{-3}$ are not shown in plots (b)-(d).}
  		\label{fig:rhosFull}
	\end{figure*}

\section{V. Bootstrapping method for reconstruction precision estimation}
	
\MODIF{
We implement a bootstrapping approach to estimate the reconstruction precision of the quantum superposition phase $\phi_a$ as a function of the sample size (number of different realizations $R$ used for the reconstruction). First, we perform $18000$ realizations with 40 resonant atomic samples, similar to those used for Fig.~2(d). The set of computed effect matrices $\{E^{(r)}\}$ is randomly split into independent groups of size $\R$ ($100\le\R\le 9000$). Using each group separately, we reconstruct $\rhoML$, determine the estimate $\phi$ of $\phi_a$ and compute its error bar $\sigma(\phi)$. With all groups of the same size $\R$, we calculate the standard deviation $\tilde\sigma_{\phi,\R}$ of the reconstructed values of $\phi$ and the mean value $\langle\sigma(\phi)\rangle_\R$ of the computed error bar~$\sigma(\phi)$. This procedure is repeated 4 times with different group samplings and the results are finally averaged.}